\documentclass[twocolumn,5p,times]{elsarticle}

\usepackage{amssymb}

\usepackage[usenames]{color}

\journal{Astroparticle Physics}

\begin{document}

\begin{frontmatter}

\title{Improving photon-hadron discrimination based on cosmic ray surface detector data}

\author[ALCALA]{G. Ros}
\ead{german.ros@uah.es}
\author[IAFE]{A. D. Supanitsky}
\ead{supanitsky@iafe.uba.ar}
\author[UNAM]{G. A. Medina-Tanco}
\ead{gmtanco@nucleares.unam.mx}
\author[ALCALA]{L. del Peral}
\author[ALCALA]{M. D. Rodr\'iguez-Fr\'ias}
\address[ALCALA]{Space and Astroparticle Group, Dpto. F\'isica y Matem\'aticas, Universidad de Alcal\'a
Ctra. Madrid-Barcelona km. 33. Alcal\'a de Henares, E-28871 (Spain).}
\address[IAFE]{Instituto de Astronom\'ia y F\'isica del Espacio, IAFE, CONICET-UBA, Argentina}
\address[UNAM]{Instituto de Ciencias Nucleares, UNAM, Circuito Exteriror S/N, Ciudad Universitaria,
M\'exico D. F. 04510, M\'exico.}

\begin{abstract}

The search for photons at EeV energies and beyond has considerable astrophysical interest and will remain one of the key challenges 
for ultra-high energy cosmic ray (UHECR) observatories in the near future. Several upper limits to the photon flux have been established 
since no photon has been unambiguously observed up to now. An improvement in the reconstruction efficiency of the photon showers and/or 
better discrimination tools are needed to improve these limits apart from an increase in statistics. Following this direction, we analyze 
in this work the ability of the surface parameter $S_b$, originally proposed for hadron discrimination, for photon search.

Semi-analytical and numerical studies are performed in order to optimize $S_b$ for the discrimination of photons from  a proton background 
in the energy range from $10^{18.5}$ to $10^{19.6}$ eV. Although not shown explicitly, the same analysis has been performed for Fe nuclei 
and the corresponding results are discussed when appropriate. The effects of different array geometries and the underestimation of the muon 
component in the shower simulations are analyzed, as well as the $S_b$ dependence on primary energy and zenith angle. 
\end{abstract}


\end{frontmatter}

\section{Introduction}
\label{intro}

Photons at EeV energies and higher are thought to be typically produced as decay secondaries in our cosmological neighborhood. They come from higher-energy 
cosmic rays (nucleon or nucleus) that interact with matter or background photons producing neutral pions and neutrons. A typical case is the Greisen, 
Zatsepin and Kuzmin (GZK) process (see e.g. Ref. \cite{Gelmini:08}) where a proton above $E_{GZK}\backsimeq60$ EeV interacts with the cosmic microwave 
background (CMB) photons losing energy and, in the most probable case, producing a neutral pion that almost immediately decay into 2 photons of about $10\%$ 
each of the initial proton energy. Neutrons could also be produced in the GZK interaction with $\sim80\%$ of the initial energy and later decay producing 
an electron and a new proton with around $10$ and $90\%$ of the neutron energy respectively. If the initial proton energy is $\gtrsim10^{20} eV$, the secondary 
electron could finally produce a photon of EeV energies through inverse Compton. Also, if UHE photons are generated in cosmologically distant sources, 
the flux is expected to steepen above the energy threshold of the GZK process since their attenuation length is only of the order of a few Mpc at such high energies.

The AGASA Collaboration on the other hand, reported a flux of UHECRs with no apparent steepening above $E_{GZK}$ \cite{AGASA:94}. Motivated by these measurements,
many theoretical models were proposed that are able to create particles of the observed energy at relatively close distances from the Earth. These models 
involve super heavy dark matter (SHDM), topological defects, neutrino interactions with the relic neutrino background (Z-bursts), etc. These are called 
\textit{top-down} models since the UHE particle is a consequence of the decay or annihilation of a more energetic entity (see Ref. \cite{Sigl:00} for a 
review). A key signature of these models is a substantial photon flux at the highest energies. Thus, the search for UHE photons was highly stimulated.
Recently, the suppression in the spectrum has been confirmed by Auger \cite{AugerSpectrum:10} and HiRes \cite{HiResSpectrum:08}, but its 
origin is still unknown and compatible with a subdominant contribution of these top-down models.

The present status is that no observation of photons has been claimed above $10^{18}$ eV by any experiment. The main candidates reported by both older 
experiments, like AGASA \cite{AGASAPhotonLimit:02} and Yakutsk \cite{YakutskPhotonLimit:10}, or the newer Pierre Auger Observatory (Auger hereafter) 
\cite{AugerPhotonSDFlux:08} and Telescope Array (T.A.) \cite{TAPhoton:11}, are all compatible with the expected fluctuations of a pure sample of very deep 
proton shower events. The most stringent upper limits to the photon flux have been established by Auger ($0.4\%$, $0.5\%$, $1.0\%$, $2.6\%$, 
$8.9\%$ for energy above $1$, $2$, $3$, $5$, $10$ EeV using hybrid data \cite{AugerPhotonICRC2011} and, $2.0\%$, $5.1\%$, $31\%$ for energy above 
$10$, $20$, $40$ EeV using surface data \cite{AugerPhotonSDFlux:08}) .

Despite the fact that no photons have been unambiguously identified up to now, a relatively small fraction of photons in the primary flux cannot be ruled 
out, and their detection would have profound implications in our understanding of the nature and origin of UHECRs. In fact, recent upper limits in the photon 
fraction constrain SHDM models in such a way that cosmic rays originated in these scenarios could only contribute in a subdominant way to the total flux. In 
addition, these limits are close to the predicted photon flux caused by the GZK interaction in certain models, whose detection would support the extragalactic 
origin of UHECRs and bring independent clues on their composition (see Ref. \cite{AlvarezMuniz:12} for a review). Also, more stringent limits on EeV photons 
reduce corresponding systematic uncertainties in the reconstruction of the energy spectrum \cite{Busca:06} and the derivation of the proton-air cross-section 
\cite{Ulrich:12}, and affect the interpretation of the observed elongation rate \cite{Unger:10}. 

Auger and the Telescope Array are the experiments that can currently detect EeV photons. Both are hybrid observatories with a ground array of detectors and 
fluorescence telescopes. At these energies, cosmic rays interact with Earth's atmosphere producing extensive air showers (EAS). EAS initiated by photon 
primaries are expected to develop deeper in the atmosphere compared to hadrons, producing larger values of $X_{max}$, the maximum of shower development 
measurable by the fluorescence telescopes. On the other hand, the surface detector exploits the fact that photon showers are characterized by a smaller number 
of secondary muons and a more compact footprint at ground. Several observables have been applied to surface data, mainly related with the spatial and temporal 
structure of the shower front at ground \cite{AugerPhotonSDFlux:08, TAPhoton:11}. A new surface parameter, called $S_b$, was proposed for proton-iron 
discrimination in Ref. \cite{Ros:11}. It is sensitive to the combined effects of the different muon and electromagnetic components on the lateral distribution 
function. In this work, we optimize $S_b$ for photon searches and analyze its specific properties for photon primaries.

The energy calibration with the surface detector is different for hadron and photon primaries, so the calculation of an upper photon limit from pure 
surface information is a complex issue. The interpolated signal at a certain distance to the shower axis is used as energy estimator ($S_{1000}$ in Auger 
\cite{AugerSpectrum:10} and $S_{800}$ in Telescope Array \cite{TASpectrum:12}) for both primaries but, comparing hadron and photon showers of the same primary energy 
and zenith angle, the difference in the energy estimator is about a factor of $2$ above $10^{18.5}$ eV, on average. Therefore, while the energy calibration for hadron 
primaries is done by using hybrid events, i.e. events seen by the fluorescence telescopes and the surface detectors simultaneously, pure Monte Carlo (MC) methods are 
used in case of photon-induced showers (see Ref. \cite{AugerPhotonSDFlux:08, Billoir:07} for Auger and Ref. \cite{TAPhoton:11} for T.A.). This energy scale difference 
is unavoidable for surface detector alone since it is a consequence of the different physics involved in hadron and pure electromagnetic showers. An unbiased measurement 
of the energy is possible if only hybrid events are used, since the primary energy is directly obtained from the longitudinal profile measured by the fluorescence 
telescopes. We assume here that the primary energy is the one used to simulate the showers (MC energy) since the problem of the different energy scales for pure surface 
events is beyond the scope of this work.

\section{Semi-analytical calculation} \label{Sec:SemiAnalytical}

In this section an improved version of the semi-analytical calculation developed in Ref. \cite{Ros:11} is 
introduced, in order to more deeply understand the behavior of the $S_b$ parameter.  

The parameter $S_b$ \cite{Ros:11}, is defined as, 
\begin{equation}
\label{Sb}
S_b = \sum_{i=1}^{N}  s_{i} \times \left(\frac{r_i}{r_0}\right)^b 
\end{equation}
where the sum extends over all triggered stations \emph{N}, $r_0$ is a reference distance ($1000$ m in the case of Auger), $s_{i}$ is 
the signal measured in the $i$th station, and $r_{i}$ is the distance of this station to the shower axis. 

The discrimination power between protons ($p$) and photons ($\gamma$) of the parameter $S_b$ can be estimated 
by using a merit factor defined as,
\begin{equation}  
\label{Eta}
\eta = \frac{E[S_b^{p}]-E[S_b^{\gamma}]}{\sqrt{Var[S_b^{p}]+Var[S_b^{\gamma}]}},
\end{equation}
where $E[S_b^A]$ and $Var[S_b^A]$ are the mean value and the variance of $S_b^A$, respectively, with $A=p, \gamma$. 

\begin{figure}[th]
\centering
\includegraphics[width=8.1cm]{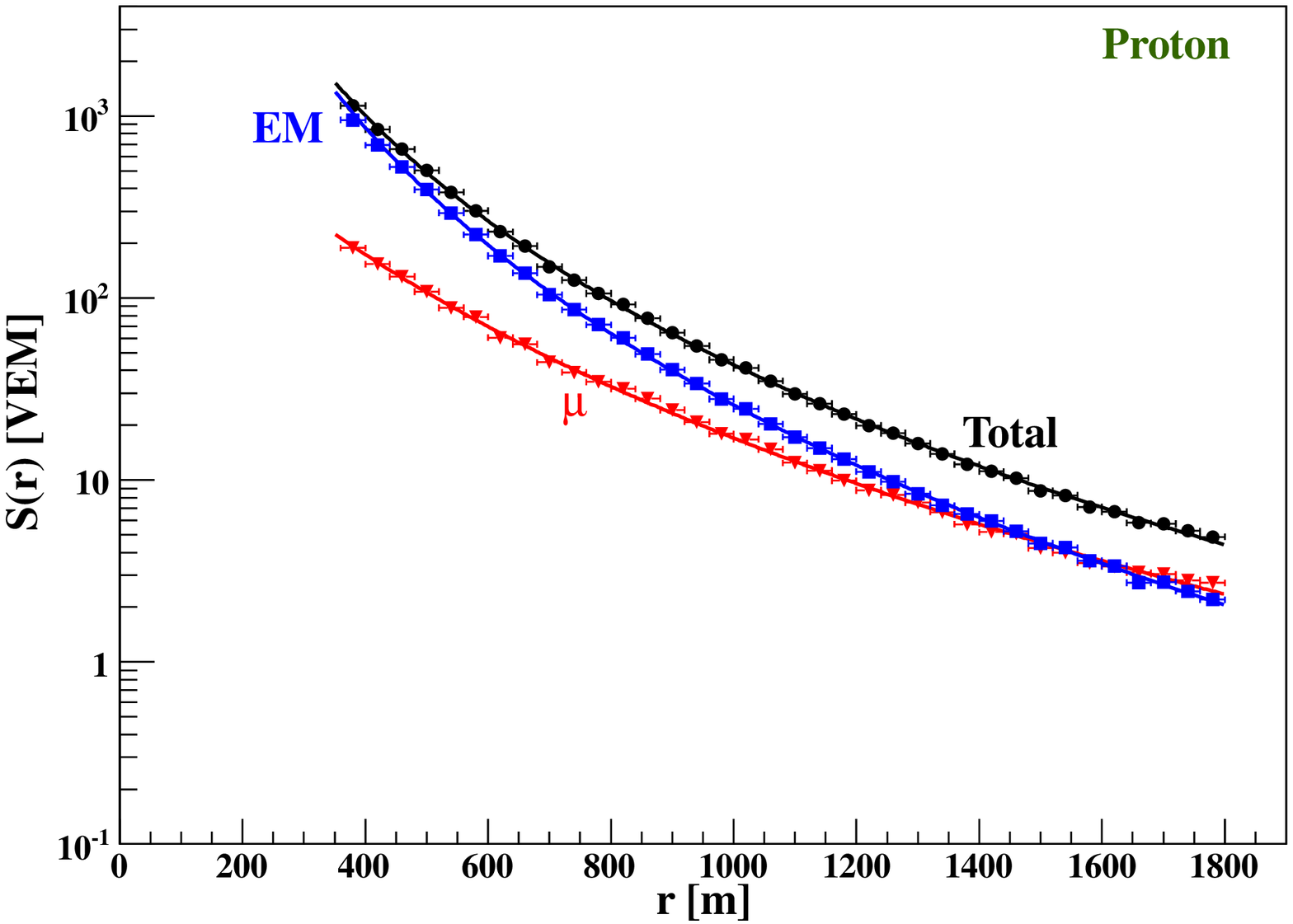}
\includegraphics[width=8.1cm]{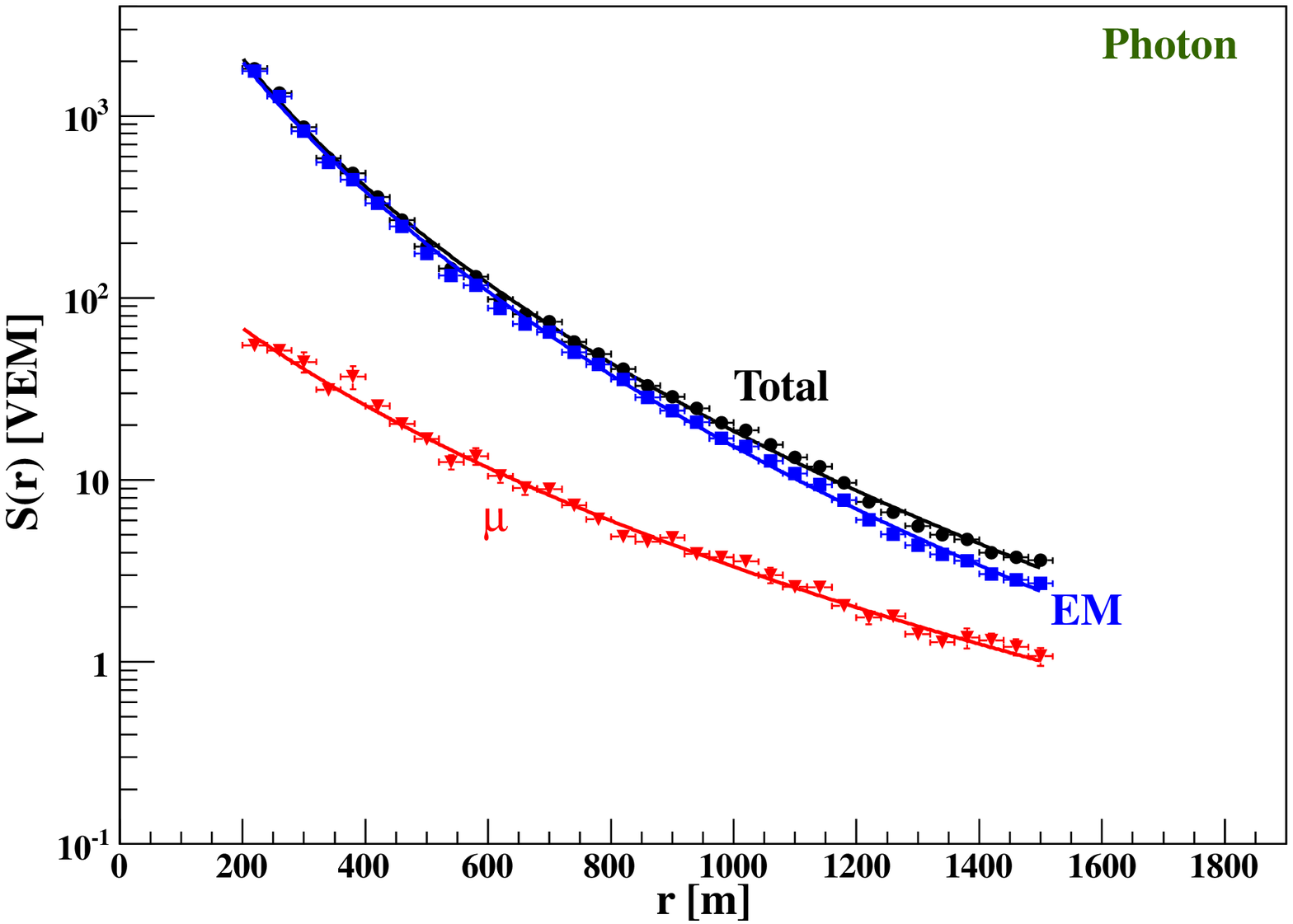}
\caption{Signal (measured in units of the energy deposited by a vertical muon, VEM) as a function of the distance to the 
shower axis for proton and photon showers obtained from simulations. The primary energy is in the interval $[10^{19}, 10^{19.1}]$ eV 
and the zenith angle is such that $1 \leq \sec\theta \leq 1.25$. Solid lines are fits to the simulated data with a NKG-like function (see text).
The hadronic interaction model used to generate the showers is QGSJET-II \cite{QGII}. 
\label{Ldfs}}
\end{figure}

The calculation of the merit factor of $S_b$ corresponding to protons and photons, by using a semi-analytical 
approach, requires the knowledge of the lateral distribution function (LDF), the signal as a function of the distance 
to the shower axis, for both protons and photons. Figure \ref{Ldfs} shows the LDFs, obtained from simulations of 
the showers impinging on Auger water Cherenkov surface detectors (see section \ref{Simulations} for details), corresponding 
to proton and photon primaries of energy in the interval $[10^{19}, 10^{19.1}]$ eV and zenith angle $\theta$, such 
that $1 \leq \sec\theta \leq 1.25$, i.e. $\theta \in [0^\circ, 36.87^\circ]$. Also shown are the LDFs corresponding to muons 
and to the electromagnetic particles (mainly electrons, positrons and photons). Solid lines correspond to the fits of the simulated 
data with a NKG-like function \cite{NKG},
\begin{equation} 
\label{NKGldf}
S(r)=S_0 \left( \frac{r}{r_0} \right)^\beta \left( \frac{r+r_s}{r_0+r_s} \right)^\alpha, 
\end{equation}
where $r_s=700$ m and $r_0=1000$ m, and $S_0$, $\beta$ and $\alpha$ are free fit parameters. For the fits of the
LDFs corresponding to the total and electromagnetic signal, the condition $\alpha=\beta$ is used, i.e. $\alpha$ is
considered as a free parameter just for the fit corresponding to the muon signal.

As expected, from figure \ref{Ldfs} it can be seen that the muon component of the photon showers is much smaller 
than the corresponding one to protons.

Following Ref. \cite{Ros:11} the distribution function for a given configuration of distances to the shower axis and signals 
(in a given event) can be written as,
\begin{eqnarray}
\label{Dist}
P(s_1, \ldots ,s_N; r_1, \ldots ,r_N) \! \! &=& \! \! f(r_1,\ldots,r_N) \times \\ \nonumber 
&& \prod^{N}_{i=1} \exp \left( -S(r_i) \right)\  \frac{S(r_i)^{s_i}}{s_i !},
\end{eqnarray}
where $r_i$ is the distance to the shower axis of the $i$th station (the first station, $r_1$, is the closest one) and 
$S(r_i)$ is the average LDF evaluated at $r_i$. Note that, in this case, the Gaussian distribution corresponding to the 
deposited signal in each station used in Ref. \cite{Ros:11} is replaced by a Poissonian distribution which is more 
suitable for small values of the total signal. Here $f(r_1,\ldots,r_N)$ is the distribution function of the random
variables $r_i$ with $i=1\dots N$, which depends on the incident flux and the geometry of the array.

From the definition of $S_b$ and Eq. (\ref{Dist}) the following expressions for the expectation value and the variance
of $S_b$ are obtained,
\begin{eqnarray}
\label{Mean}
E[S_b] \! \! \! \! \! \! &=& \! \! \! \! \! \!\sum_{i=1}^{N} E\left[ f_{E}(S(r_i)) \left( \frac{r_i}{r_0} \right)^b  \right]_r \\
\label{Var}
Var[S_b] \! \! \! \! \! \! &=& \! \! \! \! \! \! \sum_{i=1}^{N} E\left[ \left( f_V(S(r_i))-f_{E}^2(S(r_i)) \right) %
\left( \frac{r_i}{r_0} \right)^{2 b} \right]_r+ \nonumber \\
&& \! \! \! \! \! \! \sum_{i=1}^{N} \sum_{j=1}^{N} cov\left[f_{E}(S(r_i)) \left( \frac{r_i}{r_0} \right)^b, f_{E}(S(r_j)) %
\left( \frac{r_j}{r_0} \right)^b \right]_r \! \! \!,
\end{eqnarray}
where 
\begin{eqnarray}
\label{Int}
E\left[ g(r_i) \right]_r &=& \int dr_i\ g(r_i) f_i(r_i), \\
E\left[ h(r_i,r_j) \right]_r &=& \int dr_i dr_j\ h(r_i,r_j) f_{ij}(r_i,r_j),
\end{eqnarray}
see Ref. \cite{Ros:11} for details. Here $f_E(S(r_i))$ and $f_V(S(r_i))$ correspond to the mean value of $s_i$ and $s_i^2$
respectively,
\begin{eqnarray}
\label{fEfV}
f_E(S(r_i)) &=& \exp \left( -S(r_i) \right)\ \sum_{s_i=s_{min}}^{s_{max}} s_i\ \frac{S(r_i)^{s_i}}{s_i !},\\  
f_V(S(r_i)) &=& \exp \left( -S(r_i) \right)\ \sum_{s_i=s_{min}}^{s_{max}} s_i^2\ \frac{S(r_i)^{s_i}}{s_i !},
\end{eqnarray}
where it is assumed that the stations included in the $S_b$ calculation are such that $s_{min} \leq s_i \leq s_{max}$,
where $s_{min}$ corresponds to a trigger condition and $s_{max}$ to a saturation level. Taking $s_{min}=3$ VEM and assuming that
for $s_i\geq s_{max}$ the Poissonian distribution can be approximated by a Gaussian, the following expressions are obtained,
\begin{eqnarray}
\label{fEfV}
f_E(x) \! \! \! \! \! &=& \! \! \! \! \! x-\exp(-x) (x+x^2)-\sqrt{\frac{x}{2 \pi}} \times \nonumber \\  
&& \! \! \! \! \! \exp\left( -\frac{(x-s_{max})^2}{2 x} \right) -\frac{1}{2} x \left(1+\textrm{Erf}%
\left( \frac{x-s_{max}}{\sqrt{2 x}} \right) \right) \\
f_V(x) \! \! \! \! \! &=& \! \! \! \! \! x+x^2 -\exp(-x) (x+2 x^2)-\sqrt{\frac{x}{2 \pi}} \times \nonumber \\
&& \! \! \! \! \! (x+s_{max}) \exp\left( -\frac{(x-s_{max})^2}{2 x} \right) - \frac{1}{2} x (1+x) \times \nonumber \\
&& \! \! \! \! \! \left(1+\textrm{Erf}\left( \frac{x-s_{max}}{\sqrt{2 x}} \right) \right),
\end{eqnarray}
where
\begin{equation}
\textrm{Erf}(x) = \frac{2}{\sqrt{\pi}} \int_{0}^{x} dt\ \exp\left( -t^2 \right).
\end{equation}
Following Ref. \cite{Ros:09} it is assumed that $s_{max}=1221$ VEM.

The calculation of the expectation value and the variance of $S_b$ for proton and photon primaries requires the 
knowledge of the distribution function $f(r_1,\dots,r_N)$ which is very difficult to obtain analytically. Therefore, 
a very simple Monte Carlo simulation is used instead. A triangular grid of $1500$ m of distance between detectors, 
like the one corresponding to Auger, is first considered. The impact points are distributed uniformly in the central 
triangle of the array and the arrival directions of the primaries are simulated following an isotropic flux such 
that $1 \leq \sec\theta \leq 1.25$. 

\begin{figure}[th]
\centering
\includegraphics[width=8.1cm]{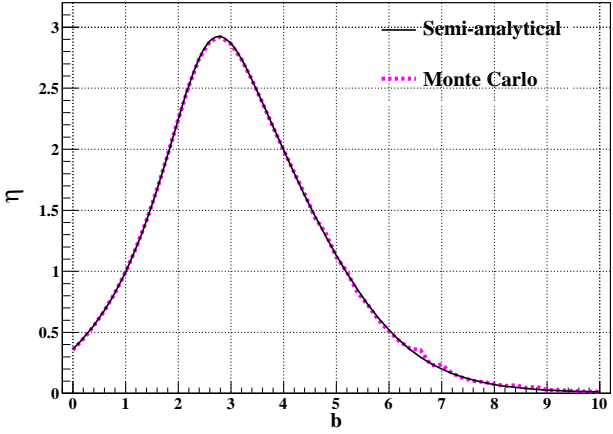}
\caption{\label{SemiMC} $\eta$ as a function of $b$ obtained by using the semi-analytical approach (solid line) 
and a simplified Monte Carlo simulation (dotted line).}
\end{figure}

The merit factor $\eta$ is calculated from Eqs. (\ref{Eta},\ref{Mean},\ref{Var}), the fitted proton and photon LDFs 
and the position of the stations obtained from the Monte Carlo simulations. Figure \ref{SemiMC} shows the comparison 
between the merit factor $\eta$ as a function of $b$, obtained by using the semi-analytical approach and a simplified 
Monte Carlo simulation, proposed in Ref. \cite{Ros:09} and also tested in Ref. \cite{Ros:11}, which includes the 
simulation of the impact points of the showers, the arrival direction and also the Poissonian fluctuations of the signal 
in each station. Note that the proton and photon LDFs used in both calculations are the same. From the figure, it can be 
seen that, as expected, $\eta$ as a function of $b$ obtained from the two different methods are in very good agreement. 
Also note that the maximum value of $\eta$ is obtained for $b\cong2.8$, very close to $b=3$.        

\subsection{Influence of fluctuations on the discrimination power of $S_b$} 

The discrimination power of $S_b$ is dominated by two type of fluctuations, the ones corresponding to 
the distance of the stations to the shower axis, which come from the uniform distribution of the impact
points of the showers over the array area, and the ones originated by the detection of the particles that
reach a given station, i.e. signal fluctuations. 

The semi-analytical approach allow us to isolate the contributions of the different sources of fluctuations that 
generate the maximum of the curve of $\eta$ as a function of $b$. Let us consider the case in which we freeze
a realization of the spatial distributions of the stations with respect to the shower core position, then  
Eqs. (\ref{Mean},\ref{Var}) become,
\begin{eqnarray}
\label{MeanS}
E[S_b] \! \! \! \! \! \! &=& \! \! \! \! \! \!\sum_{i=1}^{N} f_{E}(S(E[r_i])) \left( \frac{E[r_i]}{r_0} \right)^b, \\
\label{VarS}
Var[S_b] \! \! \! \! \! \! &=& \! \! \! \! \! \! \sum_{i=1}^{N} \left( f_V(S(E[r_i]))-f_{E}^2(S(E[r_i])) \right) %
\left( \frac{E[r_i]}{r_0} \right)^{2 b} \! \!,
\end{eqnarray}
where $E[r_i]$ is the expectation value of the distance to the shower axis of the $i$th station. Line labeled as (a)
of figure \ref{Int} corresponds to $\eta$ as a function of $b$ calculated under this approximation. It can be seen that
$\eta$ decreases for increasing values of $b$. The signal corresponding to the stations that are far from the shower 
axis presents larger fluctuations, therefore, when $b$ increases, the weight of these stations also increases making
$\eta$ to decrease.  
\begin{figure}[th]
\centering
\includegraphics[width=8.1cm]{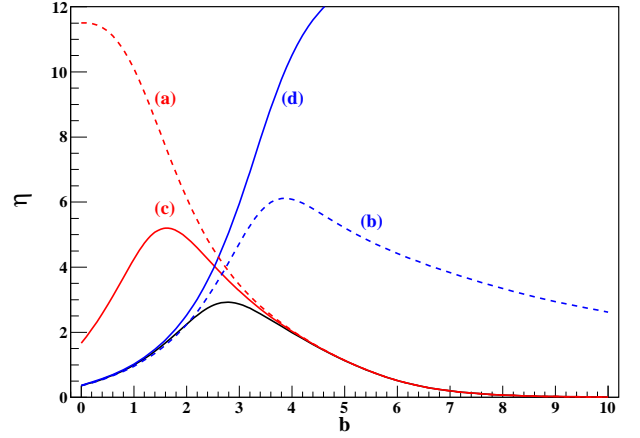}
\caption{$\eta$ as a function of $b$ in the semi-analytical approach. Black line: all contributions to the merit factor are
included (Eqs. (\ref{Mean},\ref{Var}), same curve as Fig. \ref{SemiMC}). Line (a): the signal fluctuations are only considered
(Eqs. (\ref{MeanS},\ref{VarS})). Line (b): the fluctuations in the position of the stations are only included (Eqs. (\ref{MeanR},\ref{VarR})). 
Line (c): $\eta$ calculated just considering the first term of the variance in Eq. (\ref{Var}). Line (d): as (c) but considering 
only the second term. Lines (c) and (d) include the effect of both type of fluctuations and explain the formation of the maximum 
in $\eta$ (black line). See text for more details.\label{Int}}
\end{figure}

Let us consider the other important case in which the fluctuations of the signal are switched off. In this case Eqs. 
(\ref{Mean},\ref{Var}) become, 
\begin{eqnarray}
\label{MeanR}
E[S_b] \! \! \! \! \! \! &=& \! \! \! \! \! \!\sum_{i=1}^{N} E\left[ \tilde{S}(r_i) \left( \frac{r_i}{r_0} \right)^b  \right]_r \\
\label{VarR}
Var[S_b] \! \! \! \! \! \! &=& \! \! \! \! \! \! \sum_{i=1}^{N} \sum_{j=1}^{N} cov\left[\tilde{S}(r_i) \left( \frac{r_i}{r_0} \right)^b, \tilde{S}(r_j) %
\left( \frac{r_j}{r_0} \right)^b \right]_r \! \! \!,
\end{eqnarray}
where
\begin{equation}
\label{Gamma}
\tilde{S}(r)=\left\{ 
\begin{array}{ll}
  S(r) &  \textrm{if} \ \ 3 \leq S(r)/\textrm{VEM} \leq 1221 \\
  0    &  \textrm{otherwise}
\end{array}  \right..
\end{equation}
Line labeled as (b) of figure \ref{Int} corresponds to $\eta$ as a function of $b$ calculated by using Eqs. (\ref{MeanR},\ref{VarR}). 
It can be seen that for small and for large values of $b$, $\eta$ is small. For values of $b$ close to zero the most important 
contribution to $S_b$ comes from the signal of the station closest to shower core. Therefore, due to the fast variation of the LDF with the distance 
to the shower axis, the fluctuations on the position of the first station are translated into very large fluctuations of the signal, 
decreasing drastically the discrimination power of $S_b$. The same happens for larger values of $b$ but in this case the farthest 
station is the important one. 

Note that the dominant effect for the increase of $\eta$ in the regions of $b$ where the curves (a) and (b) differ significantly 
from the exact value comes from the decrease of the variance. For the case in which the fluctuations on the positions of the stations 
are frozen the difference between the mean values is larger than the exact one for small values of $b$. However in the case where 
the signal fluctuations are frozen the difference between the mean values is smaller than the exact one for large values of $b$.     

Also note that comparing the expression of the variance for the two cases considered, Eqs. (\ref{VarS}) and (\ref{VarR}), with the 
exact expression, Eq. (\ref{Var}), it can be seen that the first term of the variance for the exact case has to do with the signal 
fluctuations and the second one with the fluctuations on the distance of the stations to the shower axis.  

Line labeled as (c) in the figure \ref{Int} corresponds to the calculation of $\eta$ in which the variance of Eq. (\ref{Var}) 
is calculated by just considering the first term. It can be seen that, for values of $b$ larger than the corresponding to the 
maximum, this term is dominated by the fluctuations of the signal. Line labeled as (d) in the figure corresponds to the 
calculation of $\eta$ in which the variance of Eq. (\ref{Var}) is calculated by just considering the second term. In this 
case it can be seen that from $b=0$ up to values close to the maximum, the behavior of $\eta$ is dominated by the fluctuations 
on the position of the stations combined with the fast variation of the LDFs with $r$. Therefore, the formation of the maximum 
in $\eta$ as a function of $b$ appears due to these two effects. Note that, the fluctuations on the position of the stations 
also contribute to the calculation of $\eta$ corresponding to line (c) and the fluctuations on the signal also contribute to 
the calculation of $\eta$ corresponding to the line (d), i.e. the exact value of the maximum cannot be obtained by just combining 
the cases in which these two kind of fluctuations are isolated.

\subsection{Modifying the muon content of showers}

\begin{figure}[th]
\centering
\includegraphics[width=8.1cm]{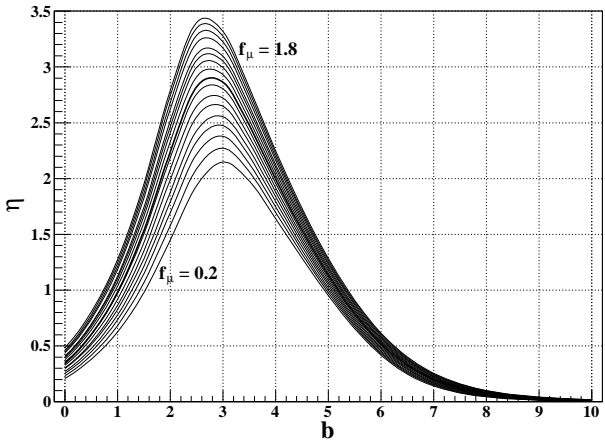}
\caption{$\eta$ as a function of $b$ for different values of $f_\mu$, ranging from $f_\mu=0.2$ to $f_\mu=1.8$ in steps of 
$\Delta f_\mu=0.1$. $f_\mu=1$ corresponds to the prediction of QGSJET-II-03. \label{Fmu} }
\end{figure}

There is experimental evidence about a deficit in the muon content of the simulated showers \cite{Engel:07,Castellina:09,Allen:11}. 
The hadronic interaction models at the highest energies cannot completely describe the observations. Therefore, the muon content 
of the showers is modified artificially, in order to study its influence on the discrimination power of $S_b$. For that purpose, the 
LDFs corresponding to the total signal, for both protons and photons, are obtained combining the fits of the LDFs corresponding to the 
electromagnetic and muon components (see figure \ref{Ldfs}) in such a way that, $S(r)=S_{em}(r)+f_\mu\ S_\mu(r)$, where $f_\mu=1$ 
corresponds to the prediction of QGSJET-II. Figure \ref{Fmu} shows $\eta$ as a function of $b$ for different values of $f_\mu$, 
from $f_\mu=0.2$ to $f_\mu=1.8$ in steps of $\Delta f_\mu=0.1$. It can be seen that the maximum value reached by $\eta$ increases 
with $f_\mu$. This is due to the fact that the difference between the mean value of $S_b$ for protons and the corresponding 
one to photons increases with $f_\mu$, as in the case of proton and iron primaries (see Ref. \cite{Ros:11} for details). Also, 
when $f_\mu$ increases the total signal increases, reducing the fluctuations of the $S_b$ parameter. Note that, $b_{opt}$, the 
value that maximize $\eta$ decreases with $f_\mu$ going from $\sim \! 3$ for $f_\mu = 0.2$ to $\sim \! 2.6$ for $f_\mu = 1.8$.       

\section{Shower and detector simulations}
\label{sim}

In this Section detector simulations are performed in order to analyze the most relevant properties of the $S_b$ parameter. 
For the calculation of $S_b$, at least $3$ triggered stations in the event are needed to assure the geometrical 
reconstruction of the shower axis. Therefore, the efficiency, i.e. the fraction of events that fulfills this requirement, is almost $100\%$
above the energy threshold of the corresponding array, highlighting a major advantage of the $S_b$ parameter. In a real experiment no quality cut
on $S_b$ is needed  except that it could be convenient to require a minimum number of active (not necessarily triggered) detectors during 
the event (for example $\geq 4$ were imposed in Ref. \cite{AugerPhotonICRC2011}) or to examine individually the few events selected as photon
candidates to avoid a possible underestimation of $S_b$ due to a missing or non-operating station which would mimic the behavior of a primary 
photon.

\subsection{$S_b$ optimization for different array sizes and geometries}
\label{Sb_Optimization_arrays}

The detection of the extensive air showers by a surface array of water Cherenkov tanks is here simulated by using our own 
simulation program described previously in Section \ref{Sec:SemiAnalytical} and Ref. \cite{Ros:09}. 
The geometry of the array and the distance  between detectors are easily modified in order to study their effect on $\eta(S_b)$. 
Thus, triangular and square grids are considered varying the array spacing from $500$ to $1750$ meters.

The error in the merit factor, $\Delta\eta$, is calculated assuming Poissonian errors and is given by,
\begin{eqnarray}  \label{EtaError}
\Delta \eta^2 = \frac{1}{Var[S_b^{p}] + Var[S_b^{\gamma}]} \times \left[ \frac{E[S_b^{p}]^2}{N_p} + \frac{E[S_b^{\gamma}]^2}{N_{\gamma}} + \right. \nonumber \\%
\left. \frac{2\eta^2}{Var[S_b^{p}]+Var[S_b^{\gamma}]} \left( \frac{Var[S_b^{p}]^2}{N_p} + \frac{Var[S_b^{\gamma}]^2}{N_{\gamma}} \right) \right], %
\end{eqnarray}
where $N_p$ and $N_\gamma$ are the number of events in each population (here $N_p = N_{\gamma} = 10^4$ are used).

\begin{figure}[th]
\centering
\includegraphics[width=8.1cm]{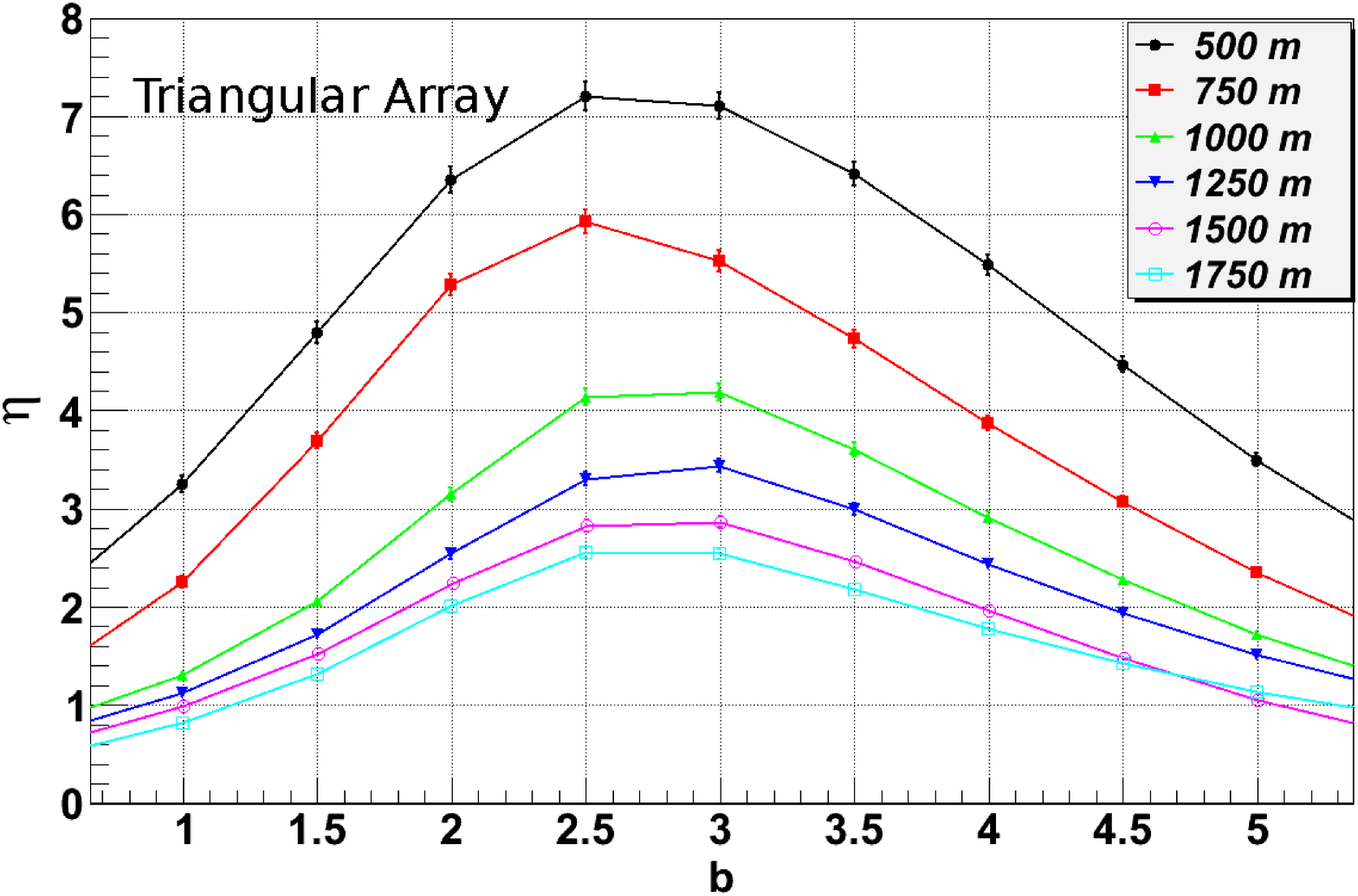}
\includegraphics[width=8.1cm]{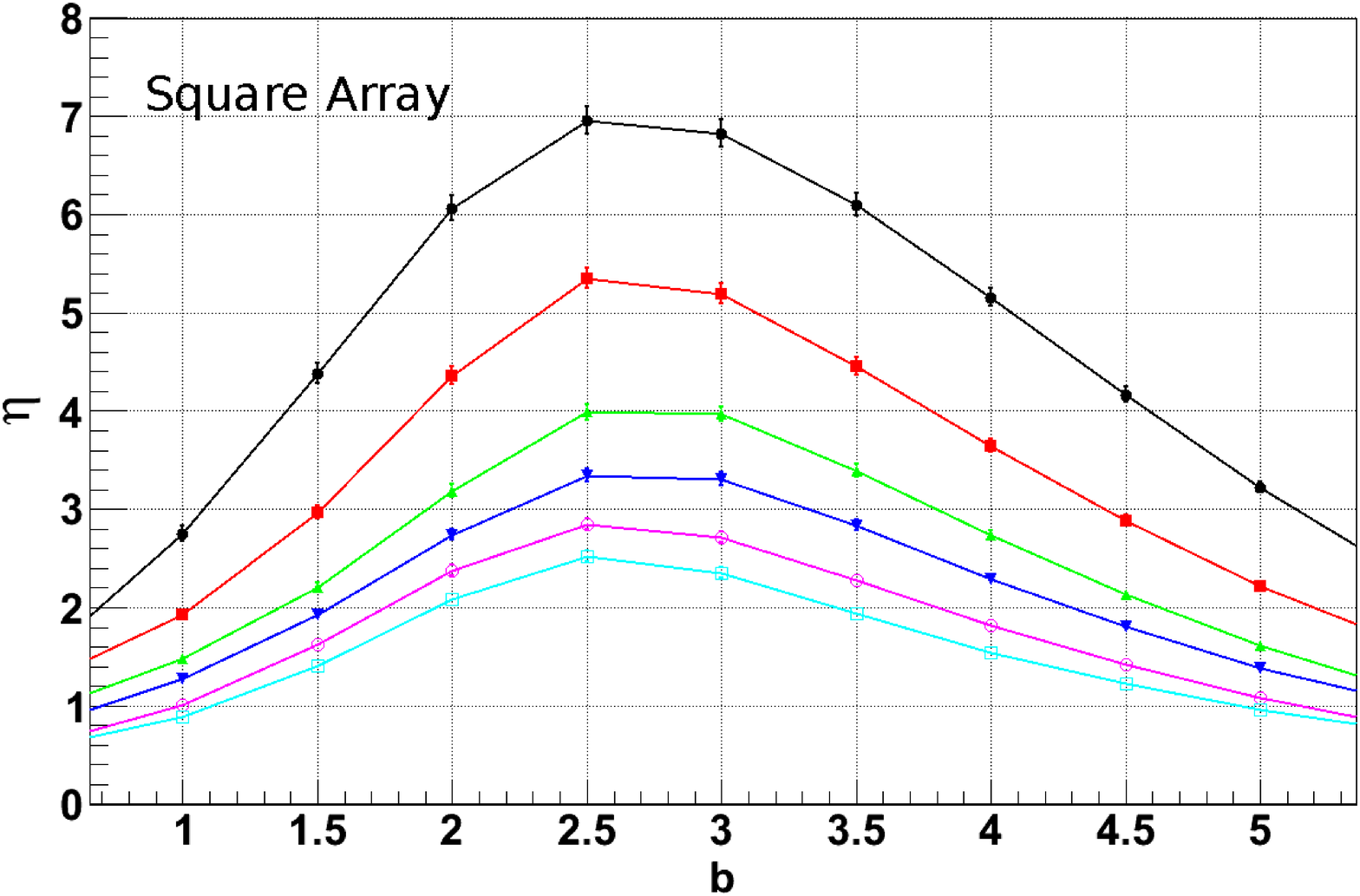}
\caption{$\eta$ as a function of $b$ for different values of the distance between detectors.}
\label{fig:MaximumMerit_vs_spacing}
\end{figure}

Figure \ref{fig:MaximumMerit_vs_spacing} shows the merit factor $\eta$ as a function of $b$ for different array sizes corresponding to 
a triangular and square grids. $\eta$ increases as the array spacing decreases as expected, since the LDF is sampled in more points 
as the array becomes denser. $\eta$ is slightly larger for the triangular grid since the number of triggered stations is also larger 
for this geometry. $b\simeq3.0$ is the optimum value for most of the arrays considered, independent of the geometry. 

\subsection{More realistic simulations} 
\label{Simulations}

In what follows, we perform a more realistic simulation in order to treat more accurately the tank response and to take into account 
the shower to shower fluctuations and experimental uncertainties such as the shower reconstruction. 

The simulation of the atmospheric showers is performed with the AIRES Monte Carlo program (version 2.8.4a) \cite{Aires} with either QGSJET-II-03 
or \cite{QGII} Sibyll 2.1 \cite{Sibyll} as the hadronic interaction model (HIM). The simulation of the tank response and the shower reconstruction 
are performed with the Offline Software provided by the Pierre Auger Collaboration \cite{AugerOffline:07}. The simulation is done for a triangular 
grid of water Cherenkov detectors of $1.5$ km of spacing, as in Auger.

The primary energy goes from $\log(E/\textrm{eV}) = 18.50$ to $19.60$ in steps of $\Delta\log(E/\textrm{eV}) = 0.05$. $1000$ events are simulated 
per each HIM and energy bin. The zenith angle follows an isotropic distribution from $0^\circ$ to $60^\circ$ while the azimuth is selected randomly 
from a uniform distribution in the interval from $0^\circ$ to $360^\circ$.

The library called MaGICS \cite{MaGICS} can be linked to AIRES in order to simulate the conversion of photons in the geomagnetic field. 
However, we do not have to deal with photon splitting, because only a negligible fraction of inclined showers convert at most latitudes
of interest below 50 EeV \cite{Homola-Conversion:07}. 

The results are very similar for both HIM, so most are only shown for QGSJET-II-03 unless otherwise stated.

\subsection{$S_b$ optimization for $\log(E/\textrm{eV})$ in $[18.5, 19.6]$ and $\theta$ in $[0^\circ, 60^\circ]$}
\label{Sb_Optimization}

\begin{figure}
\centering
\includegraphics[width=8.5cm]{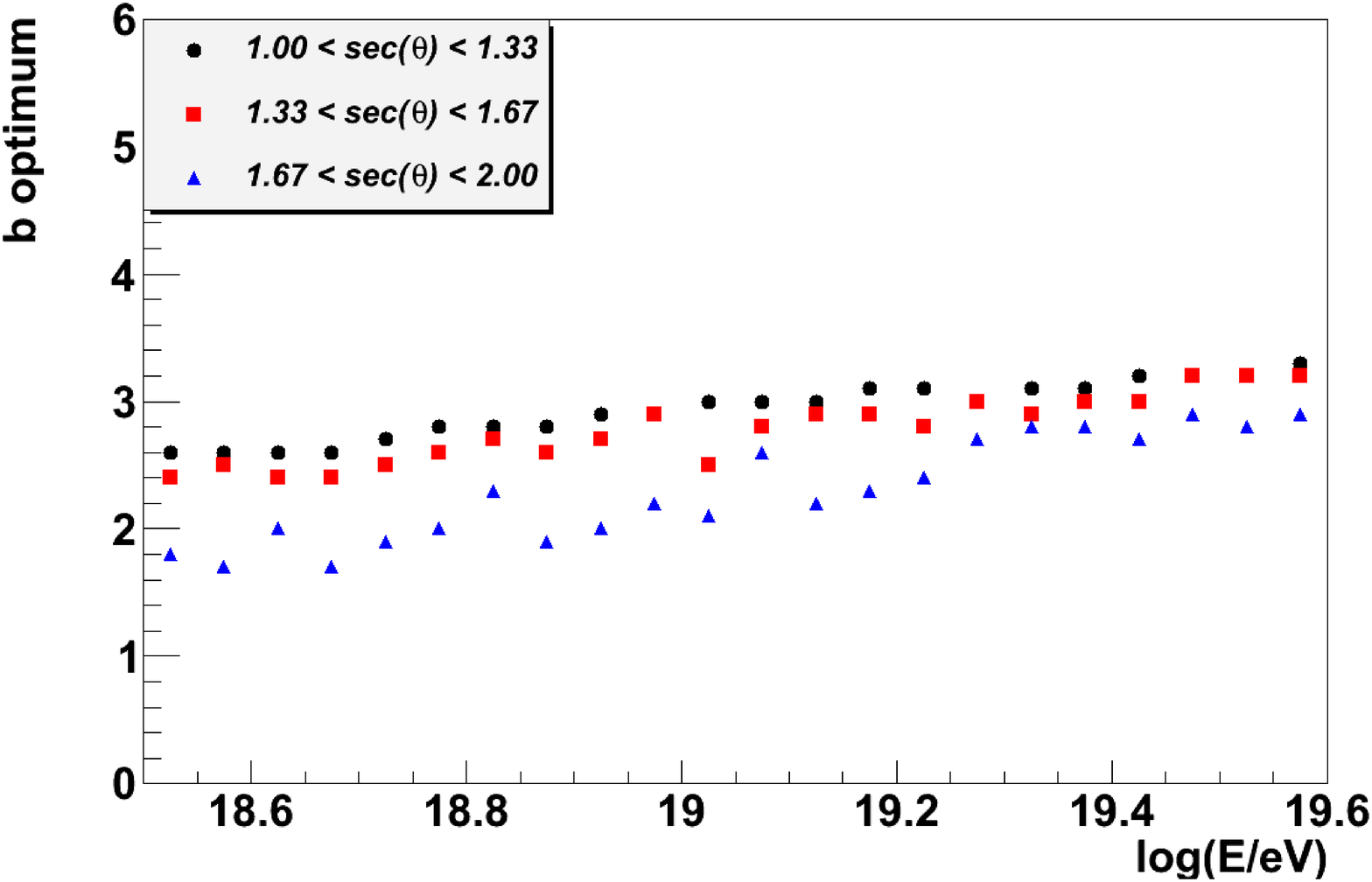}
\includegraphics[width=8.5cm]{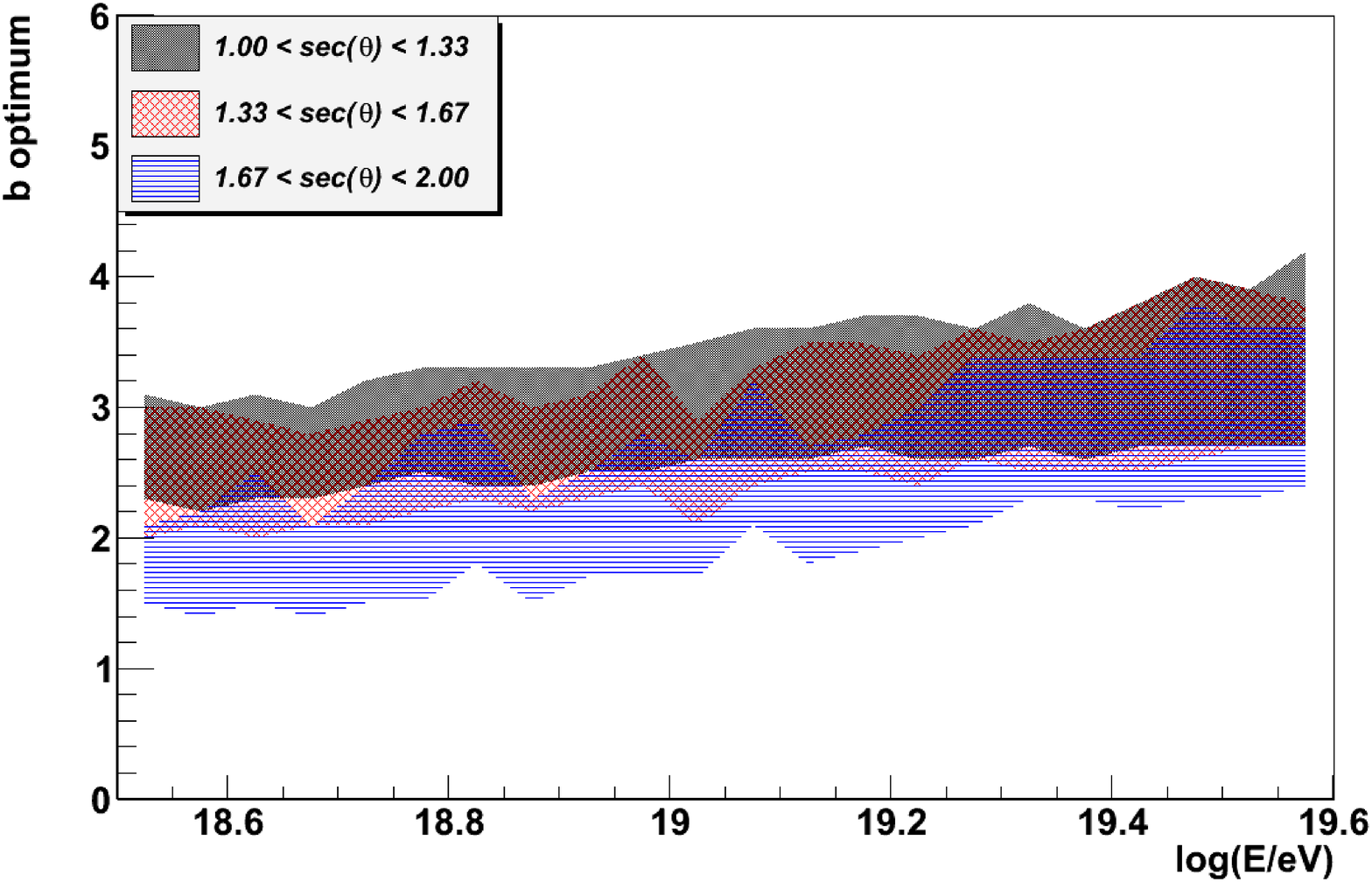}
\caption{Top: Optimum $b$ as a function of the primary energy for three different zenith angle ranges. Bottom: Bands that represent a $5\%$ variation in 
$\eta$ are added. The hadronic interaction model used is QGSJET-II-03.}
\label{fig:bopt_vs_logE}
\end{figure}

The value of $b$ that maximizes the merit factor $\eta$ as a function of the logarithm of the primary energy, $b_{opt}$, is shown in figure 
\ref{fig:bopt_vs_logE} for three zenith angle bins. In case of vertical showers with $\log(E/\textrm{eV})=19-19.1$, $b_{opt}\backsimeq3$ in agreement 
with the semi-analytical calculation (figure \ref{SemiMC}). In the bottom panel, the bands that represent a $5\%$ variation in $\eta$ are added 
showing the reliability of $S_b$ as a discriminator, even for a non-optimal selection of the index $b$.

\begin{figure}
\centering
\includegraphics[width=8.1cm]{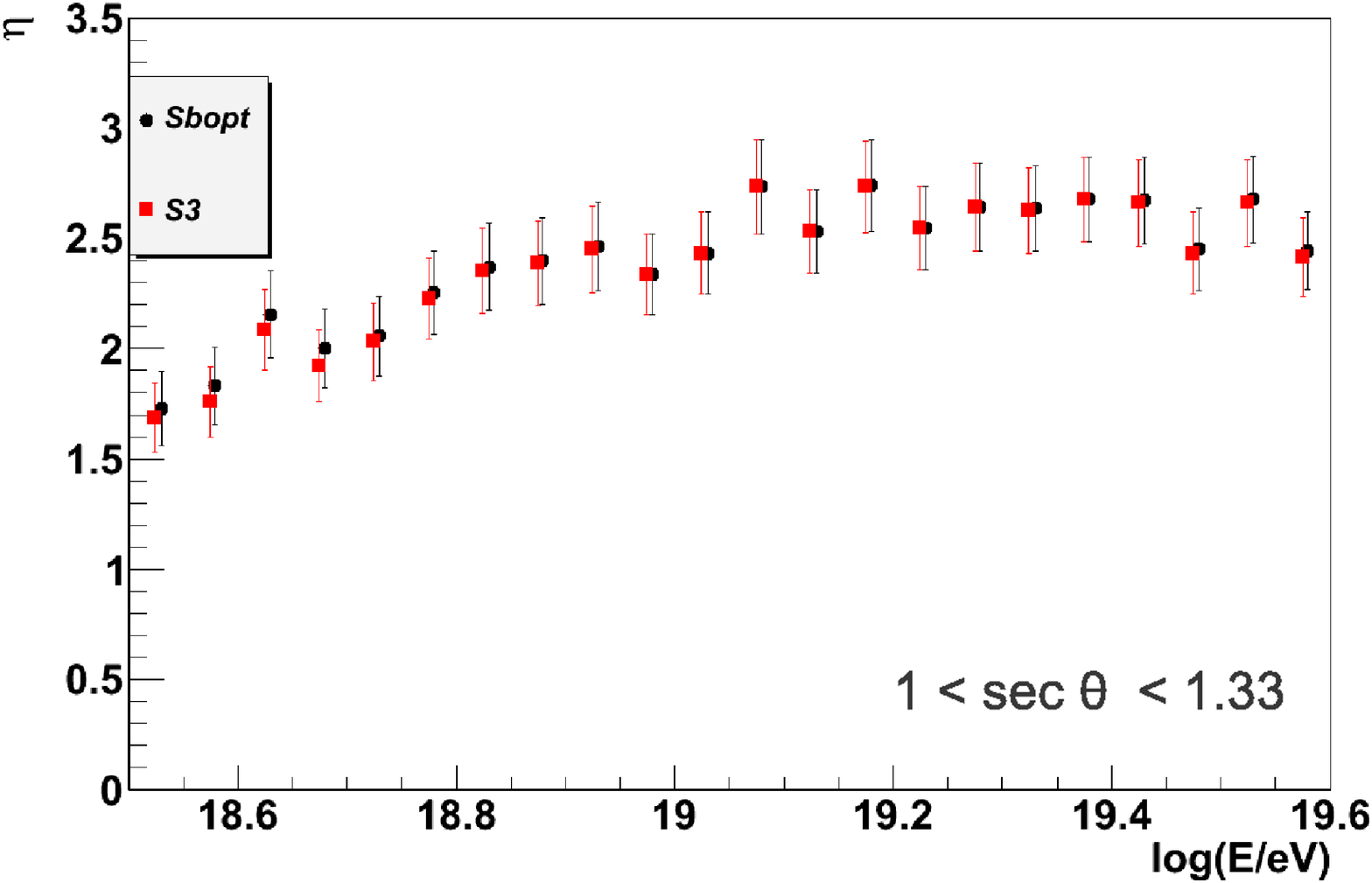}
\includegraphics[width=8.1cm]{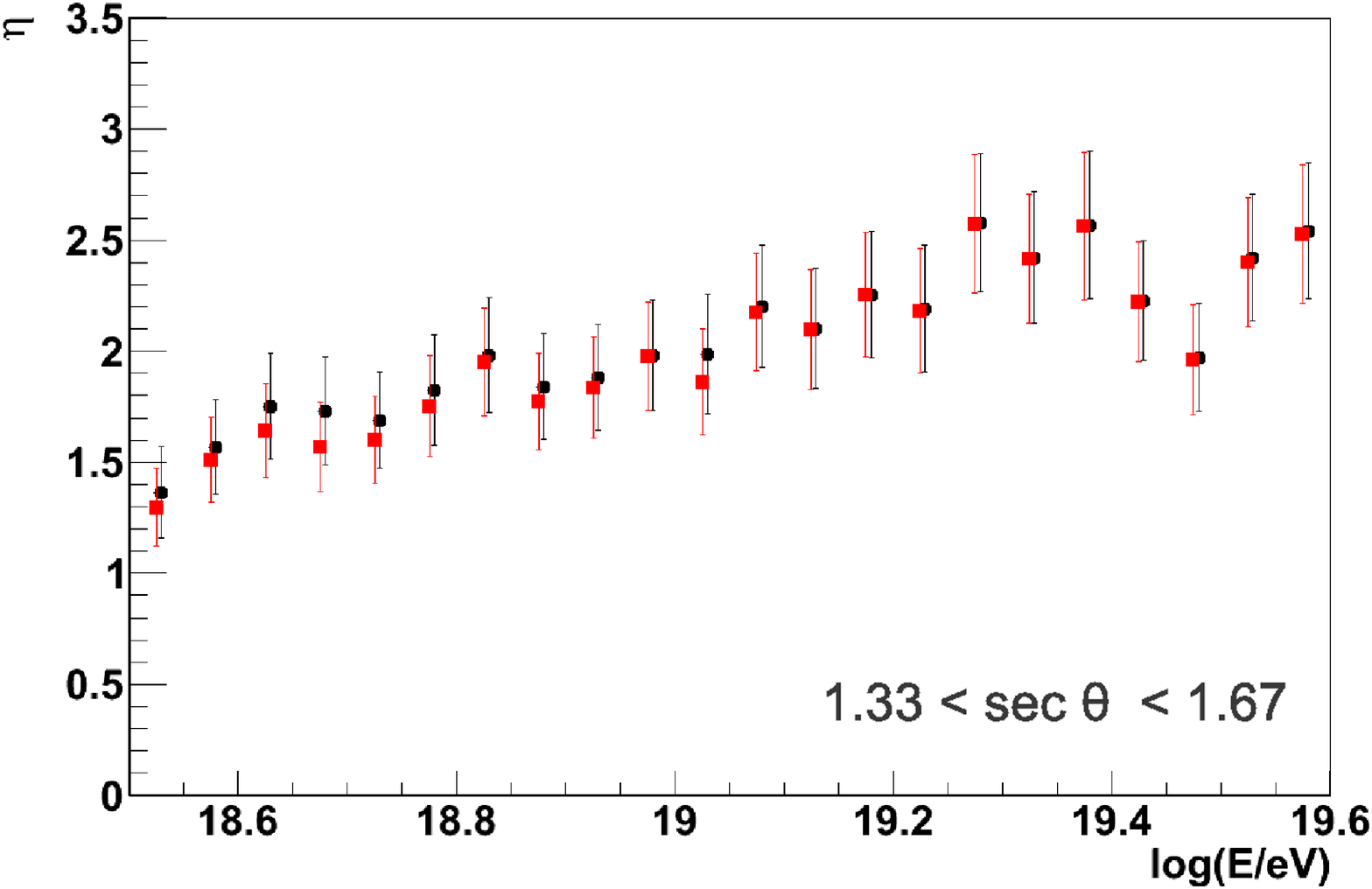}
\includegraphics[width=8.1cm]{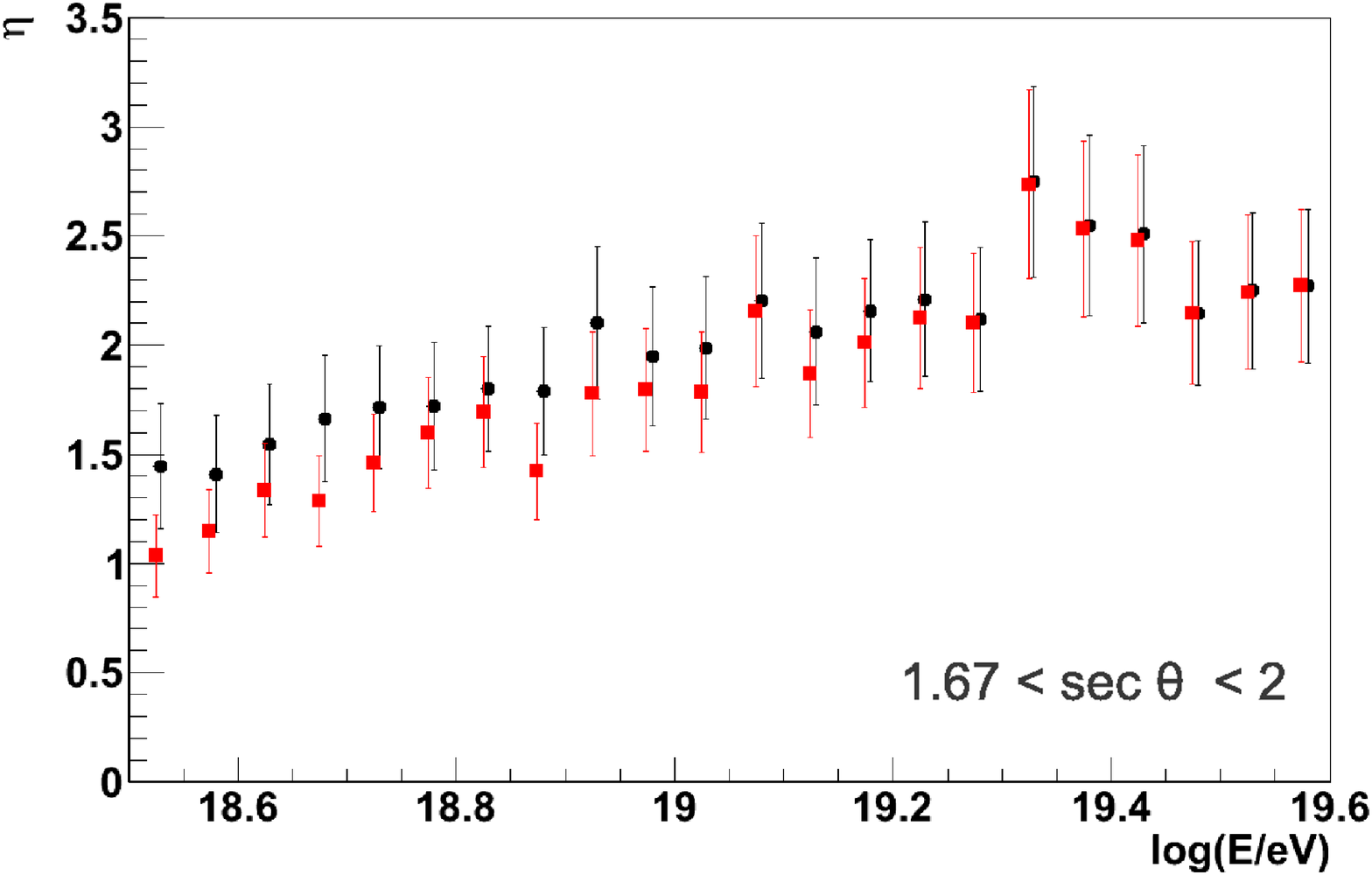}
\caption{$\eta(S_b)$ as a function of the logarithm of the primary energy for three zenith angle intervals. 
$S_b$ in case of $b=3$ and $b=bopt$ (the value that maximizes $\eta$) are shown.}
\label{fig:MF_vs_logE}
\end{figure}

From figure \ref{fig:MF_vs_logE} it can be seen that $\eta(S_3) \backsimeq \eta(S_{bopt})$ for all energies and zenith angles analyzed, except for low 
energy primaries in the small range with $sec(\theta) > 1.67$ ($\theta > 53^\circ$). Therefore, we conclude that $b=3$ is an optimum choice for the whole 
energy and zenith angle ranges analyzed, maintaining the simplicity of the parameter.

\begin{figure}
\centering
\includegraphics[width=8.5cm]{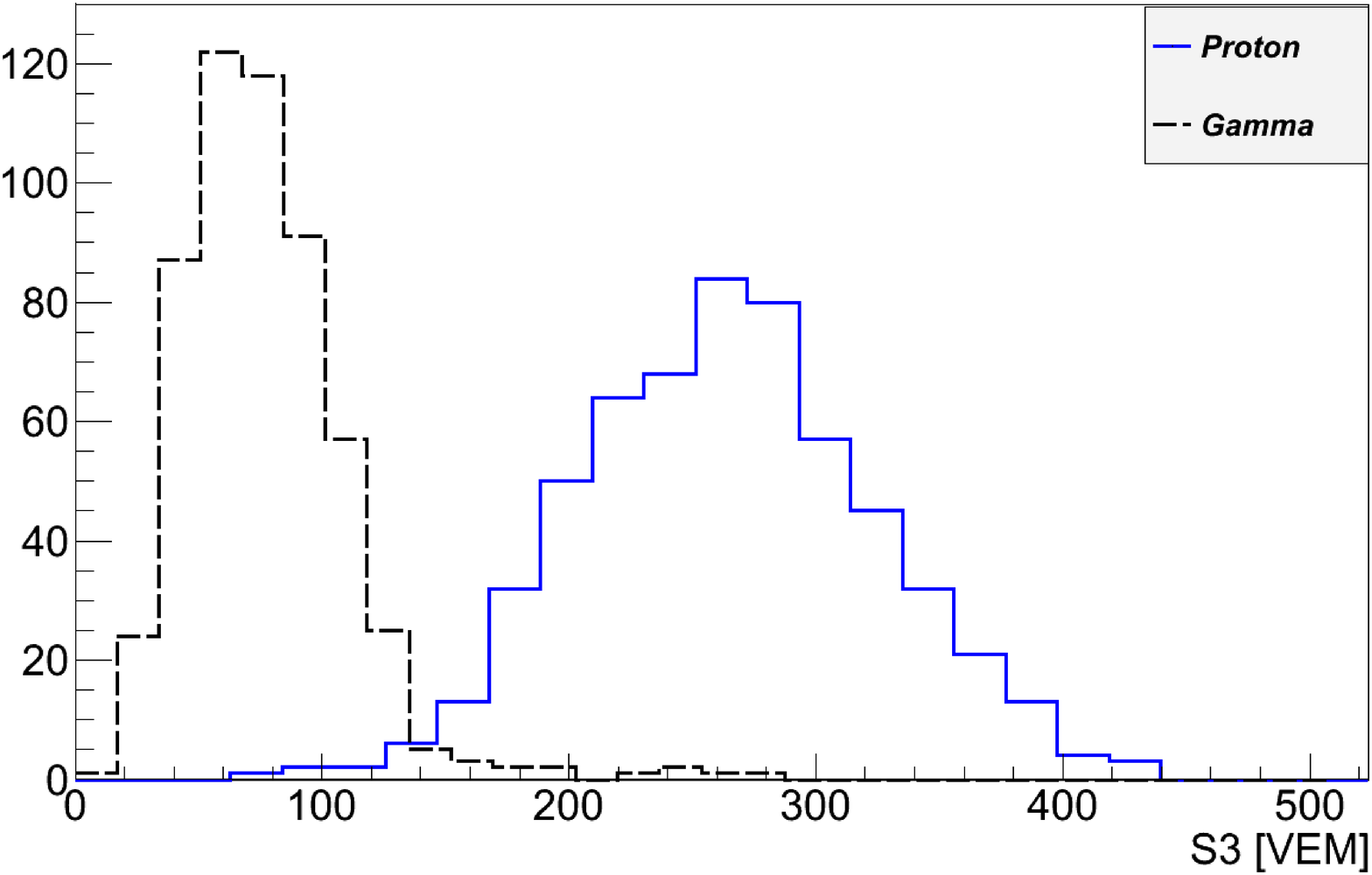}
\caption{The distribution function of $S_3$ for photon and proton initiated showers for the $1500$ m triangular array in the energy range from $\log(E/\textrm{eV})=19.05-19.10$ 
and 1.00 $< \sec(\theta) <$ 1.33.}
\label{fig:S3distributions}
\end{figure}

Although the merit factor is a good parameter to measure the statistical discrimination power of a variable, it carries by itself few information on the 
existence, shape and strength of tails of the distribution functions of the parameters. Since those tails can be also important from the point of view of 
the definition and understanding of the quality cuts, we include in figure \ref{fig:S3distributions} an example of the $S_3$ distribution functions for 
protons and photons in the energy range from $\log(E/\textrm{eV})=19.05-19.10$ and 1.00 $< \sec(\theta) <$ 1.33, where it can be seen that photon tails 
with proton-like behavior are statistically negligible but do exist.

Despite the fact that only protons have been considered so far in the analysis, a sizable fraction of heavier nuclei cannot be discarded at the highest energies 
\cite{Unger:10}. However, although not shown in this paper for brevity, equivalent calculations considering a pure iron composition show that  
$\eta(S_3)$ for photon-iron discrimination is larger than for photon-proton discrimination. Therefore, $S_{b}$, particularized for $b=3$, can be used in general for
photon-hadron discrimination with similar, or even better results, regardless of the exact UHECR mass composition.


\subsection{$S_3$ dependence with primary energy and zenith angle}
\label{Sb_energy_zenith}

\begin{figure}
\centering
\includegraphics[width=9cm]{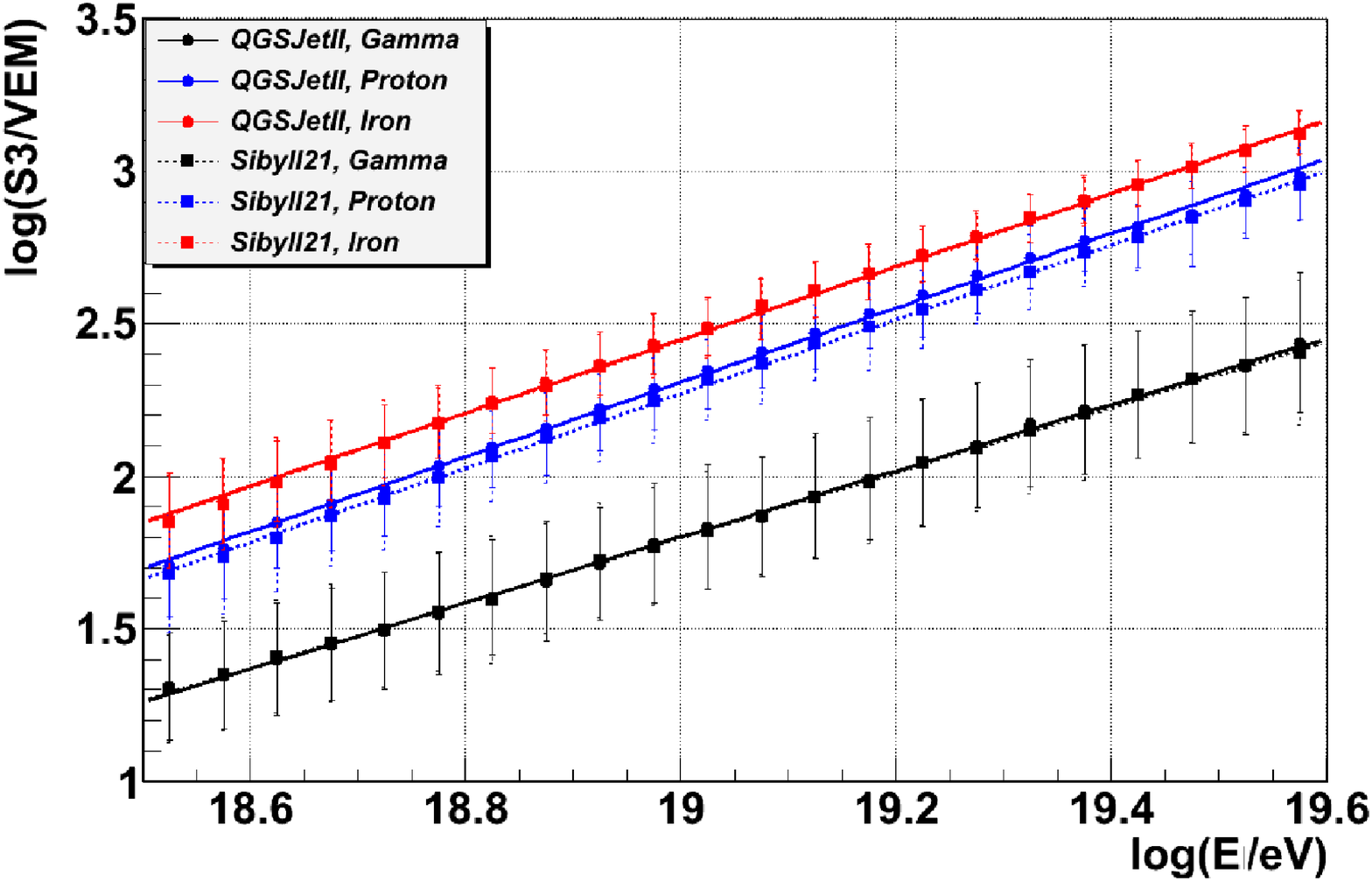}
\caption{$\log(S_3/\textrm{VEM})$ vs. $\log(E/\textrm{eV})$ for photon, proton and iron primaries. The hadronic interaction models considered 
are QGSJET-II-03 and Sibyll 2.1.}
\label{fig:S3_vs_LogE}
\end{figure}

Figure \ref{fig:S3_vs_LogE} shows the relation between $S_3$ and the primary energy. An almost linear relation is found, in agreement with Ref. \cite{Ros:11} 
where only hadrons were considered. Note that the result is almost independent of the hadronic interaction model and that the slope is smaller for photons 
compared to hadrons.

\begin{figure}
\centering
\includegraphics[width=8.1cm]{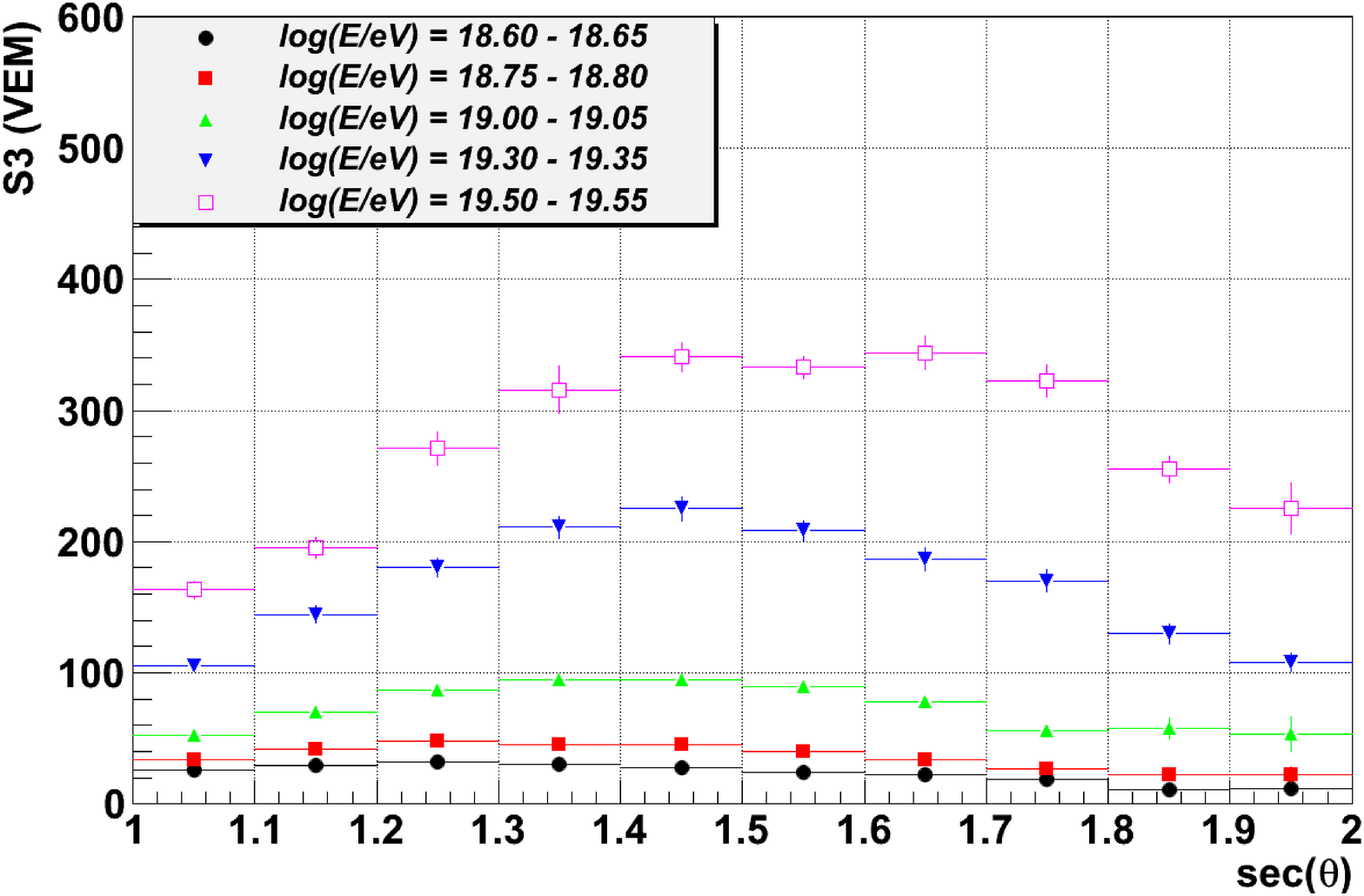}
\includegraphics[width=8.1cm]{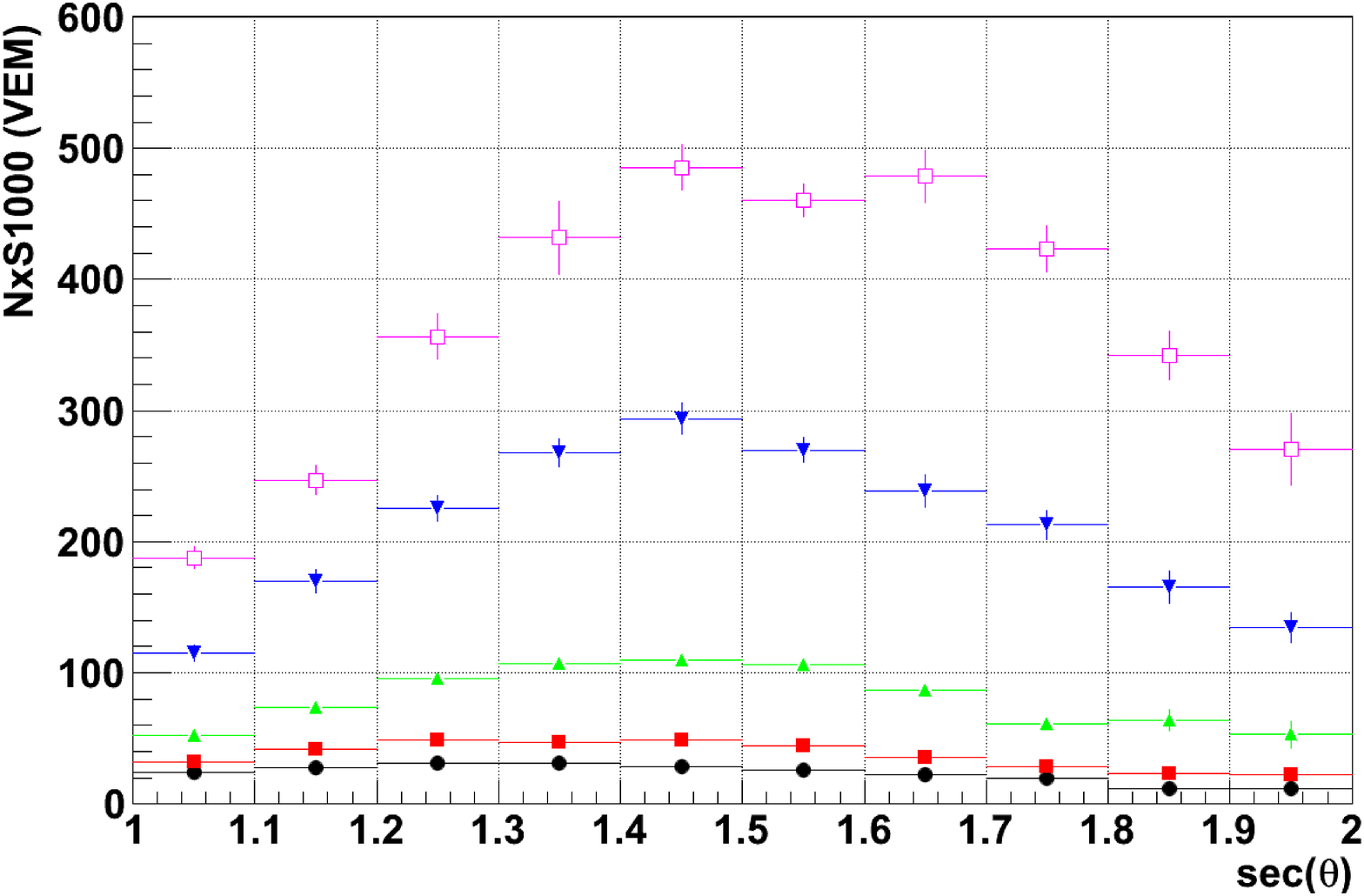}
\caption{$S_3$ (top) and $N \times S_{1000}$ (bottom) vs. $sec(\theta)$ for photon primaries and different energies. Note that the scales in the 
y-axis are the same.}
\label{fig:S3_vs_SecTheta}
\end{figure}

The dependence of $S_3$ with the zenith angle of the incoming shower for primary photons is quite complex, as shown in the top panel of figure 
\ref{fig:S3_vs_SecTheta}. While the dependence with $\sec(\theta)$ is stronger as the energy increases, the shape is similar, showing a maximum that 
slowly increases from $35^\circ$ to $50^\circ$ over a decade of energy. 

The $\theta$ dependence of $S_b$ can be qualitatively understood by considering a simplified physical situation. Let us assume that the LDF 
follows a power-law, $S(r)=S_{1000}\left(\frac{r}{r_0}\right)^{-\beta}$, where $r_0=1000$ m and $\beta$ is the slope. If $b=\beta$, then 
$S_b = N \times S_{1000}$, where $N$ is the number of candidate stations. The dependence of $N \times S_{1000}$ with zenith angle is shown in the 
bottom panel of figure \ref{fig:S3_vs_SecTheta}. $N$ is expected to increase with $\theta$ since the shower footprint at ground becomes larger and 
more elongated. On the other hand, $S_{1000}$ decreases with $\theta$ due to the larger attenuation in the atmosphere. The combination of these two 
effects roughly explain the existence of this maximum. 

\begin{figure}
\centering
\includegraphics[width=8.1cm]{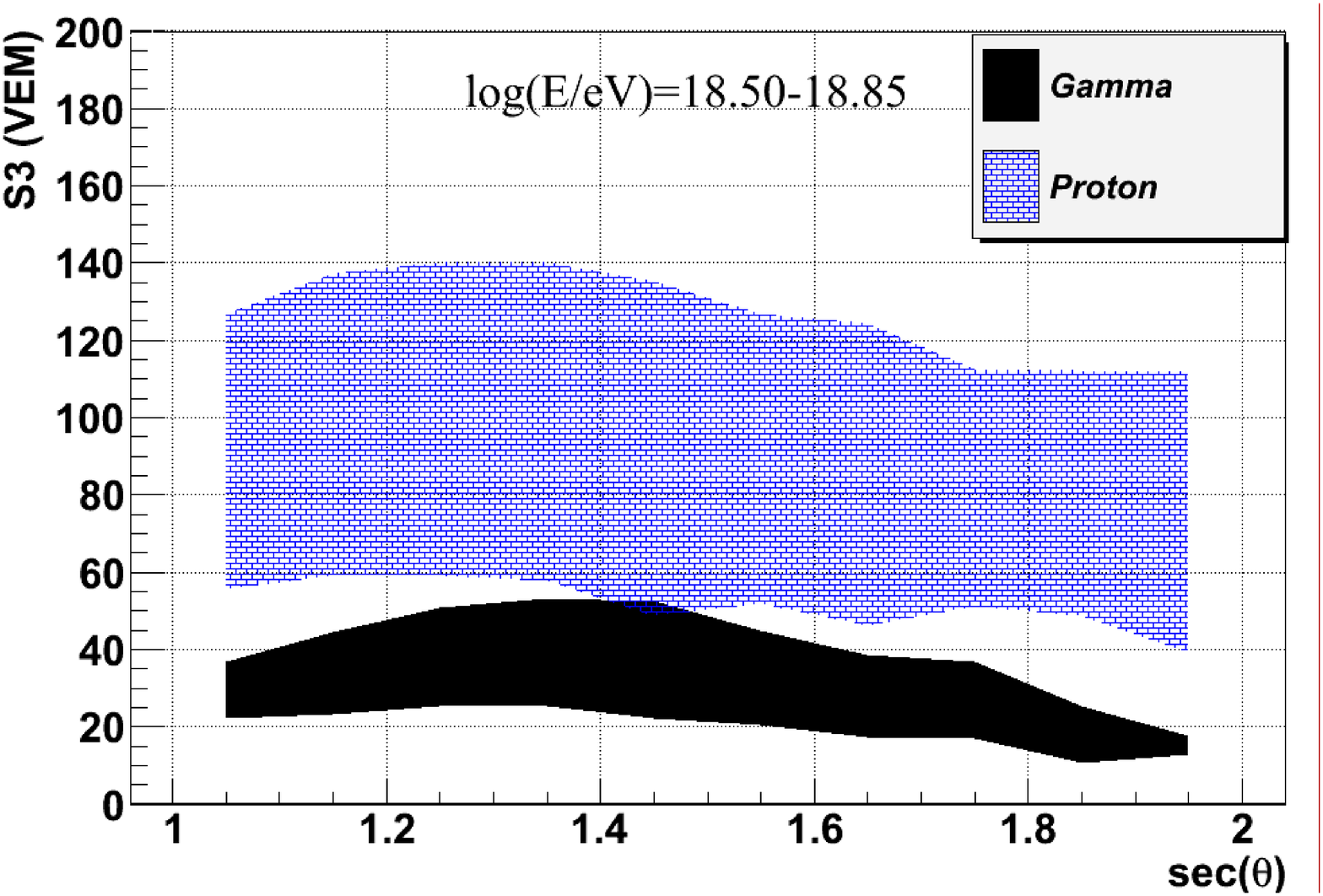}
\includegraphics[width=8.1cm]{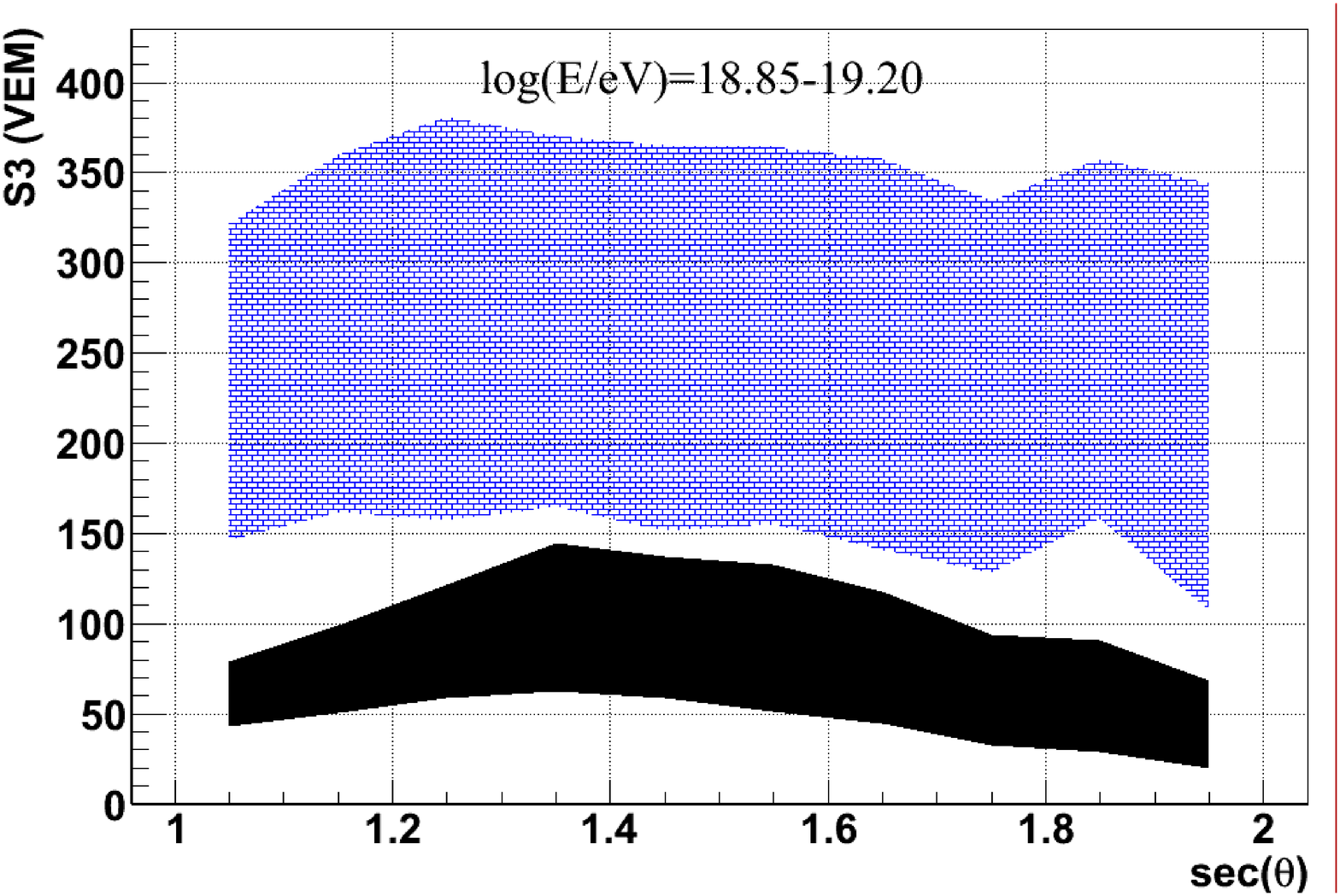}
\includegraphics[width=8.1cm]{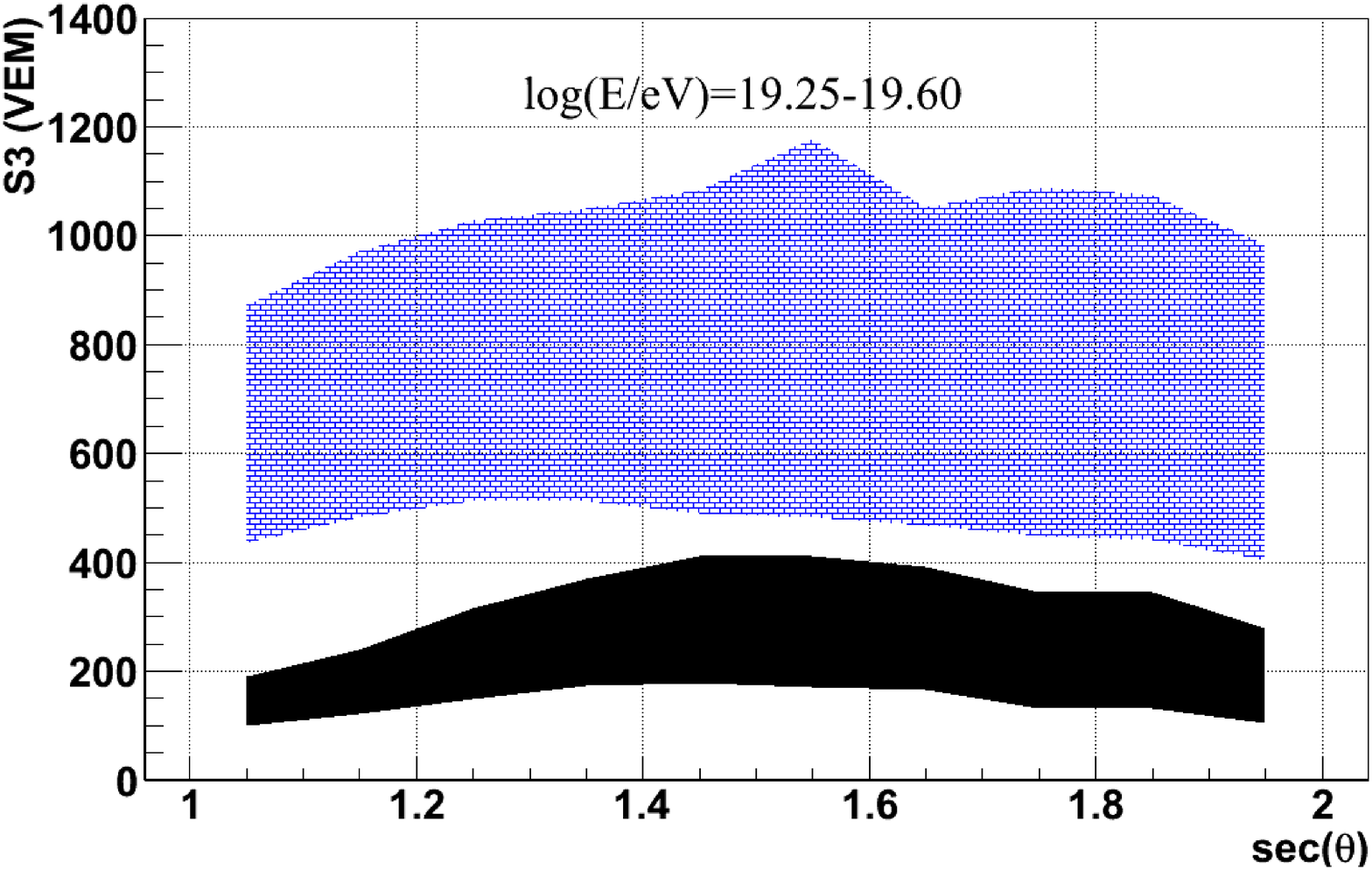}
\caption{$S_3$ vs. $sec(\theta)$ for photon and proton primaries in $3$ different energy intervals. The bands correspond to an energy interval of $\Delta \log(E/eV)=0.35$. 
Note that there is almost no overlap between both primaries.}
\label{fig:S3_gamma_proton}
\end{figure}

In the case of hadrons, $S_b$ has in general a small dependence on zenith angle, which is more manifest for quasi vertical showers at the lowest energies
(c.f., \cite{Ros:11}). In any case, as it is shown in figure \ref{fig:S3_gamma_proton}, such a dependence does not hinder the discrimination power of the 
parameter, unless the error in energy estimate is unrealistically large ($\Delta \log(E/eV)>0.35$ or $\Delta E > 50\%$).

\section{Conclusions}

We have applied the proposed $S_b$ parameter, obtained from the information given by an array of water Cherenkov detectors, to photon-hadron 
discrimination. By means of an improved semi-analytical calculation we have shown that, as in the case of proton-iron discrimination, 
there is a well defined value of the $S_b$ exponent that maximizes its discrimination capability. We have found that at $E\cong10^{19}$ eV 
the optimum value of the exponent $b$ is $\cong3$. We have demonstrated that the fluctuations on the position of the stations, combined 
with the very fast variation of the LDFs with distance, are responsible for the decrease of the merit factor at small values of $b$. On the 
other hand, we have shown that the fluctuations of the signal measured in each station are dominant at large values of $b$, decreasing 
the merit factor in this range. Therefore, the maximum of $\eta$ is attained in the transition between these two regimes. 

Experimental data suggest an excess of muons in the showers with respect to the prediction of current hadronic interaction models. By means 
of the semi-analytical calculation we have studied the effects on the $S_b$ discrimination power when the muon content of the showers is 
modified. We have found that, the optimal value of the exponent $b$ is still close to $3$ when the muon content of the showers is modified 
and that the discrimination power of $S_{3}$ is actually enhanced when the muon content of the showers increases.

This result is generalized by using two complementary and independent approaches. First, using our own simple MC program \cite{Ros:09}
of the shower detection and reconstruction, we have demonstrated that $b\cong3$ is the value that maximizes the merit factor for many different 
arrays, varying the geometry (triangular and square unitary cells) and the distance between detectors for a large range of separations 
(from $500$ to $1750$ m). Second, using a set of full numerical simulations, with a realistic tank response 
and taking into account the shower to shower fluctuations and experimental uncertainties, we have demonstrated that $b=3$ is close to the 
optimum value in the whole energy range from $10^{18.5}$ to $10^{19.0}$ eV and zenith angles from $0^\circ$ to $60^\circ$. Furthermore,
we  have also shown that the discrimination power of $S_b$ is not significantly affected even if a suboptimal value of $b$ is used.

Additionally, since the UHECR flux likely includes a sizable fraction of heavier primaries besides protons, the same analysis has been performed 
assuming the opposite scenario, i.e. a pure iron background. The discrimination power of $S_3$ is even larger in this case, confirming the fact that $S_3$ 
can be used as a composition discriminator regardless of the exact hadron composition.

We have demonstrated that $S_3$ is almost linearly dependent on the primary energy. The zenith angle dependence for photon primaries has been qualitatively 
understood in terms of the evolution of the number of triggered stations and $S_{1000}$ with the primary zenith angle. In the case of hadrons, $S_3$ 
has in general a small dependence on zenith angle which does not hinder the discrimination power of the parameter, unless the error in energy estimate is 
unrealistically large ($\Delta \log(E/eV)>0.35$ or $\Delta E > 50\%$).

The calculation of an upper photon limit from pure surface information is a great challenge since, as commented previously, the energy reconstruction method 
introduces a composition-dependent bias. This problem could be overcome if only hybrid events are considered. Then, our results suggest that $S_b$ combined 
with fluorescence observables (mainly $X_{max}$ as in Ref. \cite{AugerPhotonICRC2011}) could improve the upper limits to the photon flux in the whole 
energy range of the experiments with a unified treatment since $S_b$ is almost full-efficient above the energy threshold of the corresponding array with a 
large discrimination power.

\section{Acknowledgments}

All the authors have greatly benefited from their participation in the Pierre Auger Collaboration and its profitable scientific 
atmosphere. Extensive numerical simulations were made possible by the use of the UNAM  super-cluster \emph{Kanbalam} and the \emph{UAH-Spas} 
cluster at the Universidad de Alcal\'a. We want to thank the Pierre Auger Collaboration for allowing us to use the Auger Offline packages 
in this work, C. Bleve and B. Zamorano for fruitful discussions and J. A. Morales de los R\'ios for the maintenance 
of the \emph{UAH-Spas} cluster. We also thank the support of the MICINN Consolider-Ingenio 2010 Programme under grant MultiDark CSD2009-00064, 
Astomadrid S2009/ESP-1496, and EPLANET FP7-PEOPLE-2009-IRSES.

This work is partially supported by Spanish Ministerio de Educaci\'on y Ciencia under the projects FPA2009-11672, Mexican PAPIIT-UNAM through 
grants IN115707-3, IN115607, IN115210 and CONACyT through grants 46999-F, 57772, CB-2007/83539. ADS is member of the Carrera del Investigador 
Cient\'ifico of CONICET, Argentina.

\end{document}